\documentclass[aoas,preprint]{imsart}
\RequirePackage[OT1]{fontenc}
\RequirePackage{amsthm,amsmath}

\usepackage{graphicx}
\usepackage{amssymb}
\usepackage{amsmath}
\usepackage{epstopdf}
\usepackage{color,soul,cancel,subfigure}
\usepackage{%setspace,
mathrsfs,cancel}

\definecolor{blue2}{RGB}{050,  050,  250}
\definecolor{cyan}{RGB}{024,  055,  155}
\definecolor{pink}{RGB}{255,  105,  180}
\definecolor{purple}{RGB}{155,  000,  250}
\definecolor{red2}{RGB}{150, 000, 000}
\definecolor{red4}{RGB}{120, 000, 000}
\definecolor{green}{RGB}{021, 122, 090}
\definecolor{green4}{RGB}{021, 122, 090}
\definecolor{green2}{RGB}{000, 122, 000}
\definecolor{yellow2}{RGB}{180, 120, 000}
\definecolor{yellow4}{RGB}{120, 100, 000}
\definecolor{navy}{RGB}{000, 000, 090}

\def\be{{\bf e}}

\def\bg{{\mbox{\boldmath$g$}}}

\def\bt{{\bf t}}

\def\bx{{\bf x}}
\def\by{{\bf y}}
\def\bz{{\bf z}}

\def\bB{{\bf B}}

\def\bV{{\bf V}}

\def\thick#1{\hbox{\rlap{$#1$}\kern0.25pt\rlap{$#1$}\kern0.25pt$#1$}}

\def\bepsilon{\boldsymbol{\epsilon}}

\def\btheta{\boldsymbol{\theta}}

\def\bmu{\boldsymbol{\mu}}
\def\bnu{\boldsymbol{\nu}}
\def\bxi{\boldsymbol{\xi}}

\def\bphi{\boldsymbol{\phi}}

\def\smbalpha{\boldsymbol{{\scriptstyle{\alpha}}}}

\def\ahat{{\widehat a}}

\def\zhat{{\widehat z}}

\def\Ghat{{\widehat G}}

\def\Yhat{{\widehat Y}}

\def\qtilde{{\widetilde q}}

\def\ttilde{{\widetilde t}}

\def\Qtilde{{\widetilde Q}}

\def\bzhat{{\widehat \bz}}

\def\lambdahat{{\widehat\lambda}}
\def\muhat{{\widehat\mu}}

\def\xihat{{\widehat\xi}}

\def\sigmahat{{\widehat\sigma}}

\def\phihat{{\widehat\phi}}

\def\Sigmahat{{\widehat\Sigma}}

\def\xitilde{{\widetilde\xi}}

\def\bthetahat{{\widehat\btheta}}

\def\bmuhat{{\widehat\bmu}}

\def\bxihat{{\widehat\bxi}}

\def\bphihat{{\widehat\bphi}}

\def\smbalpha{\widehat{\smbalpha}}

\def\bxitilde{{\widetilde\bxi}}

\def\hbar{\bar{ h}}

\def\ybar{\bar{ y}}
\def\zbar{\bar{ z}}

\def\Qbar{\bar{ Q}}

\def\Ksc{{\cal K}}
\def\Lsc{{\cal L}}

\def\Qsc{{\cal Q}}

\def\Ssc{{\cal S}}
\def\Tsc{{\cal T}}
\def\Usc{{\cal U}}

\def\transpose{{\sf \scriptscriptstyle{T}}}

\def\half{\frac{1}{2}}

\def\nhalf{n^{\half}}

\def\nnhalf{n^{-\half}}

\def\var{\mbox{var}}

\def\sumin{\sum_{i=1}^n}

\def\trans{^{\transpose}}
\def\inv{^{-1}}

\def\cov{\mbox{cov}}

\def\var{\mbox{var}}

\def\mybox#1{\vskip1mm \begin{center}
        \hspace{.0\textwidth}\vbox{\hrule\hbox{\vrule\kern6pt
\parbox{.9\textwidth}{\kern6pt#1\vskip6pt}\kern6pt\vrule}\hrule}
        \end{center} \vskip-5mm}
\def\lboxit#1{\vbox{\hrule\hbox{\vrule\kern6pt
      \vbox{\kern6pt#1\vskip6pt}\kern6pt\vrule}\hrule}}

\def\thickboxit#1{\vbox{{\hrule height 1mm}\hbox{{\vrule width 1mm}\kern6pt
          \vbox{\kern6pt#1\kern6pt}\kern6pt{\vrule width 1mm}}
               {\hrule height 1mm}}}

\def\fat#1{\hbox{\rlap{$#1$}\kern0.25pt\rlap{$#1$}\kern0.25pt$#1$}}

\def\Abb{\mathbb{A}}

\def\sumin{\sum_{i=1}^n}

\def\inv{^{-1}} 	
\def\Tsc{\mathcal T}

\def\var{\text{Var}}
\def\cov{\text{Cov}}

\def\xiik{\xi_{ik}}

\def\sm{^{(m)}}
\def\smstar{^{*(m)}}

\def\Vbb{\mathbb{V}}

\def\Usc{\mathcal{U}}
\def\Abb{\mathbb{A}}

\begin{document}
%\arxiv{math.PR/0000000}

\begin{frontmatter}
\title{Functional principal variance component testing for a genetic association study of HIV progression}
\runtitle{FPVC testing for HIV progression}

\begin{aug}
\author{\fnms{Denis} \snm{Agniel}\ead[label=e1]{denis.agniel@mail.harvard.edu}},
\author{\fnms{Wen} \snm{Xie}\ead[label=e2]{xiew06@gmail.com}},
\author{\fnms{Myron} \snm{Essex}\ead[label=e3]{messex@hsph.harvard.edu}},
%\and
\author{\fnms{Tianxi} \snm{Cai}\ead[label=e2]{tcai@hsph.harvard.edu}}

\affiliation{Harvard Medical School and Harvard T. H. Chan School of Public Health}

\address{Denis Agniel\\
Department of Biomedical Informatics\\
Harvard Medical School\\
10 Shattuck St\\
Boston, MA 02115\\
\printead{e1}}
%\phantom{E-mail:\ }\printead*{e2}}

\address{Wen Xie\\
Department of Immunology and Infectious Diseases\\
Harvard T. H. Chan School of Public Health\\
655 Huntington Ave\\
Boston, MA 02115}

\address{Myron Essex\\
Department of Immunology and Infectious Diseases\\
Harvard T. H. Chan School of Public Health\\
655 Huntington Ave\\
Boston, MA 02115}

\address{Tianxi Cai\\
Department of Biostatistics\\
Harvard T. H. Chan School of Public Health\\
655 Huntington Ave\\
Boston, MA 02115}
\end{aug}

\begin{abstract}
HIV-1C is the most prevalent subtype of HIV-1 and accounts for over half of HIV-1 infections worldwide. Host genetic influence of HIV infection has been previously studied in HIV-1B, but little attention has been paid to the more prevalent subtype C. To understand the role of host genetics in HIV-1C disease progression, we perform a study to assess the association between longitudinally collected measures of disease and more than 100,000 genetic markers located on chromosome 6. The most common approach to analyzing longitudinal data in this context is linear mixed effects models, which may be overly simplistic in this case. On the other hand, existing non-parametric methods may suffer from low power due to high degrees of freedom (DF) and may be computationally infeasible at the large scale. We propose a functional principal variance component (FPVC) testing framework which captures the nonlinearity in the CD4 and viral load with potentially low DF and is fast enough to carry out thousands or millions of times. The FPVC testing unfolds in two stages. In the first stage, we summarize the markers of disease progression according to their major patterns of variation via functional principal components analysis (FPCA). In the second stage, we employ a simple working model and variance component testing to examine the association between the summaries of disease progression and a set of single nucleotide polymorphisms. We supplement this analysis with simulation results which indicate that FPVC testing can offer large power gains over the standard linear mixed effects model.
\end{abstract}



\end{frontmatter}

\section{Introduction}
An important goal of large-scale genomic association studies
is to explore susceptibility to complex diseases. These studies have led to identification of many genomic regions as putatively harboring disease susceptibility alleles for a wide range of disorders. For patients with a particular disease, association studies have also been performed to identify genetic variants associated with progression of disease. The disease progression is often monitored by longitudinally measured biological markers. Such longitudinal measures allow researchers to more clearly characterize clinical outcomes that cannot necessarily be captured in one or even a few measurements. 
%For example, trajectories of a measure called disease activity score (DAS) are often used to quantify the progression of rheumatoid arthritis (RA) \cite{prevoo1995modified}, and forced expiratory volume can similarly be used to monitor the progression of lung health \cite{yang2009genetic}. For HIV-infected patients, the trajectories of HIV RNA (viral load) and CD4$^+$ T cell counts are important parameters for monitoring the disease and determining optimal treatment strategies.

We are motivated by a large-scale association study of HIV-1 Subtype-C (HIV-1C) progression in sub-Saharan African individuals. HIV-1C is the most prevalent subtype of HIV-1 and accounts for over half of HIV-1 infections worldwide \cite{geretti2006hiv}. Sub-Saharan Africa, where HIV-1C dominates, was home to an estimated 69\% of people living with HIV in 2012 \cite{joint2012global}. While a number of human leukoucyte antigen (HLA) alleles \cite[e.g.]{fellay2007whole, van2009association, migueles2000hla} and other loci have been identified to be associated with AIDS progression in European males infected by HIV-1B \cite{fellay2007whole, o2013host}, comparatively little research has focused on host genetic influence in this African population and subtype. 

In this study, we seek to relate the longitudinal progression of these two markers -- log$_{10}$ CD4 (lCD4) count and log$_{10}$ viral load (lVL) -- to a set of approximately 100,000 single nucleotide polymorphisms (SNPs) located on chromosome 6 in two independent cohorts of treatment-naive individuals in Botswana. We focus on chromosome 6 because it houses the HLA region of genes, which are known to impact immune function. Throughout this paper, we will use $\by = (y_1, ..., y_r)\trans$ to denote the longitudinal outcome (in the context of this study, either lCD4 or lVL), measured at times $\bt = (t_1, ..., t_r)\trans$. We will furthermore let a set of genetic markers of interest be denoted $\bz$ and any potential covariates be $\bx$. 

The most common approach to analyzing longitudinal data of this kind is to use linear mixed effects (LME) models \cite{laird1982random}, which relate $\by$ linearly to $\bz$, $\bx$, and $\bt$ with both fixed and random effects. However, both in the case of HIV progression measured by lCD4 and lVL and in general, such a linear relationship is likely to be overly simplistic. To incorporate nonlinear trajectories, many generalizations of LMEs have been proposed. Typically, methods assume that $\by$ is a noisy realization of a smooth underlying function $Y(\cdot)$. These include nonlinear or nonparametric LMEs using a fixed spline basis expansion \cite{lindstrom1990nonlinear, rice2001nonparametric,guo2002functional} for $\bt$ and adjusting for $\bz$. Functional regression methods have been proposed for densely sampled trajectories \cite{chiou2003functional} for a single $z$ as well as for irreguarly spaced longitudinal data where $\bz$ is also allowed to change over time \cite{yao2005functionalb}. In these methods, estimation proceeds via kernel smoothing for both the population mean function and the covariance process of $Y(\cdot)$. \cite{krafty2008varying} proposed an iterative procedure for fitting functional regression models which accounts for within-subject covariance but does not estimate random effects. Similarly, \cite{reiss2010fast} proposed a ridge-based estimator for the case when $\by$ is measured on a common, fine grid of points for all subjects, but requires $z$ to be univariate and also ignores random effects. In a similarly dense setting \cite{morris2006wavelet} proposed wavelet-based mixed effects models, and inference procedures for the random effects were developed in \cite{antoniadis2007estimation}.

While some of these methods can be adapted to test for association, none of them are appealing for our study for the following reasons. First, some require restrictive assumptions about the density of measurements \cite[e,g]{morris2006wavelet} which are clearly not met here. Further, all of these methods were developed with estimation and regression in mind. While many of them could in principle be used to derive testing procedures, the validity of their inference procedures often rely on the model assumptions on the distribution of $\by$ given the SNPs and the covariates. Under model mis-specification, the resulting test may fail to maintain type I error. Additionally, these procedures would require fitting a complex iterative or smoothing-based model thousands or millions of times for genome-wide studies and hence become computationally infeasible. We, on the other hand, propose a novel testing procedure which is valid regardless of the true distribution of $\by$, is fast to fit and has a simple limiting distribution, despite the fact that we account for nonlinearity in $\by$ in a manner akin to previous functional regression methods. To do this, we first capture the nonlinearity in $\by$ using functional principal components analysis (FPCA) \cite{castro1986principal,rice1991estimating,yao2005functional,hall2006properties,krafty2008varying}, Using a eigenfunction decomposition of the smoothed covariance function of $Y(\cdot)$, we approximate each patient's $Y(\cdot)$ by a weighted average of the estimated eigenfunctions, with weights corresponding to functional principal component {\em loadings} or {\em scores}. Then borrowing a variance component test framework and using these scores as pseudo outcomes, we construct a {\em Functional Principal Variance Component (FPVC)} test that can capture the nonlinear trajectories without requiring a normality assumption or fitting individual functional regression models. The test statistics can be approximated by a mixture of chi-square and the small number of eigenfunctions needed to approximate the trajectories can result in a low degree of freedom in the resulting test statistic.

Since the data of interest is sampled at irregular time intervals, we use the best linear unbiased predictor (BLUP) to estimate the scores. The BLUP was also the basis for FPCA with sparse longitudinal data in the {\em principal analysis via conditional expectation} (PACE) method \cite{yao2005functional} under a normality assumption. Here, we use BLUP to motivate our testing procedure but do not require normality for the validity of the FPVC test. The test statistic can be derived through the variance component testing framework and viewed as a summary measure of the overall covariance between the estimated subject-specific scores,  which characterize the person's trajectory, and the genetic markers. Similar variance component tests have previously been proposed for standard linear and logistic regressions with observed single outcomes \cite{wu2011rare}.

The primary virtues of FPVC testing are threefold. First, we separate the procedure into two stages of distinct complexity in order to make it feasible at large scale. In the first stage, we model $\by$ flexibly using FPCA and obtain a succinct summary of disease progression for each patient, once and for all. In the second stage, we perform a rather simple model at large scale. Thus, we segregate the computationally complex stage (which need occur only once) from the large-scale stage (which could require the same computation on the order of millions of times in, e.g., genome-wide association studies). Second, the summary of $\by$ that we obtain from FPCA is the most succinct summary possible, as the eigenfunctions identified by FPCA are the functions that explain the most variability in $\by$. Third, our theoretical results suggest that the null distribution of the FPVC test statistic reduces to a simple mixture of $\chi^2$ distributions. The variability due to estimating the eigenfunctions does not contribute to the null distribution of the test statistic asymptotically at the first order (see \eqref{u} and the derivation of the asymptotic null distribution).

The structure of the rest of the paper is as follows. In section 2, we describe FPCA and introduce FPVC testing and our main theoretical results. In section 3, we give details about the association study for HIV progression. In section 4, we discuss simulation results, and in section 5 we discuss further implications of our procedure.

\section{Functional principal variance component testing}%\vspace{-0.35in}
\subsection{The test statistic}\label{s:ts}
In this section, we propose a testing procedure for assessing the association between a set of genetic markers $\bz$ and a longitudinally measured outcome $\by$, adjusting for covariates $\bx$.
Let the data for analysis consist of $n$ independent random vectors $\Vbb = \left\lbrace \bV_i = (\by_i\trans, \bt_i\trans, \bz_i\trans, \bx_i\trans)\trans\right\rbrace_{i=1}^n$, where $\by_i = (y_{i1}, ..., y_{ir_i})\trans$ is a vector of outcome measurements taken at times $\bt_i = (t_{i1}, ..., t_{ir_i})\trans \in \Tsc^{r_i}$, $\Tsc$ is a closed and bounded interval, $\bz_i = (z_{i1}, ..., z_{ip})\trans$ is a vector of genetic markers of interest, and $\bx_i = (1, x_{i1}, ..., x_{iq})\trans$ is a vector of additional covariates that are potentially related to the outcome, all measured on person $i$. For each $i$, we take $(\bz_i\trans, \bx_i\trans)\trans$ to be distributed as $(\bz\trans, \bx\trans)\trans$.

Our goal is to test the null hypothesis
\begin{equation}
H_{0} : \quad \by_i \perp \bz_{i\Ssc} \mid \bx_i   \label{hypothesis}
\end{equation}
where $\bz_{i\Ssc} = (z_{ij_1}, ..., z_{ij_s})\trans$ is a set of genetic factors to test, identified by the index set $\Ssc = \{j_1, ..., j_s\} \subset \{1, ..., p\}$. Special cases include marginal testing, as in traditional genome-wide association studies, where $\Ssc = \{j\}$ for some $j \in \{1, ..., p\}$, or set-based testing where $\Ssc$ are the indices of the SNPs in a gene or some other related set.

To model the longitudinal trajectory, we assume that $y_{ir}$ is a noisy sample of a smooth underlying function $Y_i(\cdot)$, evaluated at the point $t_{ir}$,
\[y_{ir} = Y_i(t_{ir}) + \epsilon_{ir},\]
% where, for each $i$, $Y_i(\cdot)$ is distributed as $Y(\cdot)$ and $E\{Y(t)\}= \mu(t)$, 
% $\cov\{Y(s), Y(t)\} = G(s, t)$,  for any $s,t \in \Tsc$, and $\epsilon_{ir}$ is a random error independent of $Y_i(\cdot), \bx_i$, and $\bz_i$, with
% $E(\epsilon_{ir}) = 0$, and $\var(\epsilon_{ir}) = \sigma^2$. Furthermore, we assume there is an orthogonal expansion of $G$, \[
% G(s, t) = \sum_{k=1}^{\Ksc} \lambda_k\phi_k(s)\phi_k(t),
% \]
% for $s,t \in \Tsc$, where $\{\phi_k(\cdot)\}_{k=1}^\Ksc$ are the eigenfunctions associated with non-negative nonincreasing eigenvalues $\{\lambda_k\}_{k = 1}^\Ksc$, and $\Ksc$ could be infinity. Then we can express the observed data as a linear combination of the population mean $\mu(\cdot)$ and the eigenfunctions 
which following the logic of \cite{yao2005functional} can be written as a linear combination of its population mean $E\{Y(\cdot)\}$ and a set of $\Ksc$ eigenfunctions $\{\phi_k(\cdot)\}$
\begin{align}
y_{ir} =  \mu(t_{ir}) + \sum_{k=1}^{\Ksc} \xiik \phi_k(t_{ir}) + \epsilon_{ir}.\label{ymodel}
\end{align}
where $\xiik$ is the FPCA score associated with the $k$th eigenfunction, $E(\xiik) = 0, \var(\xiik) = \lambda_k$, and the eigenfunctions are ordered such that the $k$th explains the $k$th most variance in $Y(\cdot)$.
%$\xiik = \int_\Tsc \left\{ Y_i(t) - \mu(t)\right\}\phi_k(t)dt$ are independent random variables with $E(\xiik) = 0$ and $\var(\xiik) = \lambda_k$.
Thus, the relationship between $\by_i$ and $\bx_i$ and $\bz_i$ must be captured by the only random quantity in \eqref{ymodel}, the
random coefficients $\{\xiik\}_{k=1}^{\Ksc}$. Therefore, testing \eqref{hypothesis} is equivalent to testing
$$H_0 : \quad \{\xi_{ik}\}_{k=1}^{\Ksc} \perp \bz_{i\Ssc} \mid \bx_i.$$
However, direct assessment of the association between $\{\xi_{ik}\}_{k=1}^{\Ksc}$ and $\bz_{i\Ssc}$ is difficult
since $\{\xi_{ik}\}_{k=1}^{\Ksc}$ are unobservable and $\Ksc$ could be infinity. 

%To handle the high dimensionality in $\{\xi_{ik}\}_{k=1}^{\Ksc}$, we note that, despite the fact that the underlying trajectory $Y_i(\cdot)$ is a linear combination of a potentially infinite number of eigenfunctions, we in general need only a finite number to account for nearly all of the variability in $Y_i(\cdot)$. And because the eigenfunctions are ordered so that the $k$th eigenfunction explains the $k$th most variability in $\by$, we can then approximate $Y_i(\cdot)$ using only the first $K$ eigenfunctions
Most methods of estimating a function like $Y_i(\cdot)$ require an explicit tuning of the smoothness of the resulting estimator. Here that corresponds to choosing a (typically small) number $K$ of eigenfunctions as an approximation \[Y_i(t) \approx \mu(t) + \sum_{k=1}^K \xi_{ik} \phi_k(t),\]
where $K<\infty$ could be chosen such that the first $K$ directions capture a proportion of the variation at least as large as $\wp \in (0,1]$. Our simulation results (see section \ref{simulation}) suggest that the performance of FPVC testing is not very sensitive to the choice of $K$ provided that $\wp$ is close to 1.
Since  $\{\xi_{ik}\}_{k=1}^{K}$ are not observable or generally estimable due to sparse sampling of measurement times, we instead infer about 
the association between $Y_i(\cdot)$ and $\bz_i$ based on the {\em best linear unbiased predictor} (BLUP), 
%with observable quantities, we propose to use the so-called BLUP
\begin{align}
\xitilde_{ik} = \lambda_k\bphi_{ik}\trans\Sigma_{\by_i}\inv(\by_i - \bmu_i)\label{xitilde}
\end{align}
where $\bphi_{ik} = \{\phi_k(t_{i1}), ..., \phi_k(t_{ir_i})\}\trans, \bmu_i = \{\mu(t_{i1}), ..., \mu(t_{ir_i})\}\trans$, and $\Sigma_{\by_i} = \cov(\by_i, \by_i)$ such that $\left(\Sigma_{\by_i}\right)_{rl} = G(t_{ir}, t_{il}) + \sigma^2\delta_{rl}$, $\cov\{Y(s), Y(t)\} = G(s, t)$, and $\delta_{rl} = I_{\{r = l\}}$. In the PACE method of \cite{yao2005functional}, $\xitilde_{ik}$ was obtained as $E(\xiik | \by_i)$ under the assumption that $\xiik$ and $\epsilon_{ir}$ are jointly normal, but we don't require normality here. We simply take $\xitilde_{ik}$ as an observable and reasonable approximation to $\xiik$ even if normality does not hold, as has been argued in \cite{robinson1991blup} and \cite{jiang1998asymptotic}.

Thus, we propose to test \eqref{hypothesis} by testing
$$ H_0^\dag: \quad \{\xitilde_{ik}\}_{k=1}^{K} \perp \bz_{i\Ssc} \mid \bx_i.$$ Taking note that the association we seek to test is conditional on $\bx$, one may construct a test for $H_0^{\dag}$ by regressing $\bxitilde_i = (\xitilde_{i1}, ..., \xitilde_{iK})\trans $ onto $(\bx_i, \bz_{i\Ssc})$. However, this is only valid if
the effect of $\bx_i$ on $Y_i(\cdot)$ is captured fully based on the model relating $\bx_i$ and $\bxitilde_i$, which may not be true in general.
To remove the effect of $\bx_i$ without imposing a strong assumption on how $\bx_i$ affects $Y_i(\cdot)$,
we instead choose to model the conditional expectation of $z_{ij}$ given $\bx_i$, $\mu_{z_j}(\bx_i) = E(z_{ij} | \bx_i)$, and
center $\bz_{i\Ssc}$ as $\bz^*_{i\Ssc} = (z^*_{ij_1}, ..., z^*_{ij_s})$ where for any $j$
\[z_{ij}^* = z_{ij} - \mu_{z_j}(\bx_i).\]

To form the test statistic for $H_0$, we propose to summarize the overall association between $Y(\cdot)$ and $\bz_\Ssc$ based on the Frobenius norm of the standardized covariance between $\bxitilde_i$ and $\bz_{i\Ssc}^*$
\begin{align}\label{q0}
Q_0 =  \left\|\nnhalf\sumin \bxitilde_i \bz_{i\Ssc}^{*\transpose}\right\|_F^2.
\end{align}
Though $Q_0$ takes a simple form and can be motivated naturally as an estimated covariance (and can thus be considered model-free), it can also be viewed as a variance component score test statistic similar to those considered previously for other regression models \cite{commenges1995score,lin1997variance}.
Details on the derivation of the variance component score test statistic are given in section \ref{connections}.

Both $\bxitilde_i$ and  $\bz^*_{i\Ssc}$ involve various nuisance parameters that remain to be estimated.
First, under mild regularity conditions which are outlined in the supplementary article \cite{supp},
% outline in appendix \ref{app_fpca}
we can use FPCA to estimate 
%$\mu(\cdot), \lambda_k, \phi_k(\cdot), G(\cdot, \cdot),$ and $\sigma^2$ by
%$\muhat(\cdot), \lambdahat_k, \phihat_k(\cdot), \Ghat(\cdot, \cdot),$ and $\sigmahat^2$, respectively, 
the relevant quantities via local linear smoothing as in \cite{hall2006properties} and \cite{yao2005functional}.
Subsequently, we can estimate $\xitilde_{ik}$ by \begin{align}
\xihat_{ik} = \lambdahat_k\bphihat_{ik}\trans\Sigmahat_{\by_i}\inv(\by_i - \bmuhat_i)\label{xihat}
\end{align}
for $\bphihat_{ik} = \{\phihat_k(t_{i1}), ..., \phihat_k(t_{ir_i})\}\trans, \bmuhat_i = \{\muhat(t_{i1}), ..., \muhat(t_{ir_i})\}\trans$, and $\left(\Sigmahat_{\by_i}\right)_{rl} = \Ghat(t_{ir}, t_{il}) + \sigmahat^2\delta_{rl}$.
To estimate $\mu_{z_j}(\bx_i)$, various approaches can be taken depending on the nature of $\bx$. For example,
when $\bx$ is discrete, $\mu_{z_j}(\bx_i)$
can be estimated empirically. With continuous $\bx$, we may impose a parametric model with
\begin{equation}
\mu_{z_j}(\bx) = g_j(\btheta_j, \bx) \label{model-zw}
\end{equation}
and obtain $\zbar_j(\bx)$ as $g_j(\bthetahat_j,\bx)$, where $\bthetahat_j$ is an estimate of a
finite-dimensional parameter $\btheta_j$. To take two examples that commonly come up in genomics, if $z_j$ takes values in $\{0, 1\}$, e.g. under the dominant model, then $z_j$ can be modeled using logistic regression, and if $z_j$ takes values in $\{0,1,2\}$, then we may use a binomial generalized linear model or a proportional odds model. There are two reasons to prefer to remove the effect of $\bx$ from $\bz$ rather than from $\bxi$: first, it may in general be easier to specify a model for $\bz$ rather than $\bxi$ because of the limited range of $\bz$, and, second, this formulation facilitates asymptotic analysis without the need to derive the asymptotic distributions for the estimated FPCA scores.
Finally, based on $\{\xihat_{ik}\}_{k=1}^K$ and  $\zbar_j(\bx_i)$, our proposed test statistic is
\begin{align}\label{teststat}
Q = \frac{1}{n} \sum_{j \in \Ssc}\sum_{k=1}^K \left(\sum_{i=1}^n \xihat_{ik} \zhat^*_{ij}\right)^2 = \left\|\nnhalf\sum_{i=1}^n \bxihat_{i} \bzhat_{i\Ssc}^{*\transpose} \right\|_F^2 .
\end{align}
where $\bzhat_{i\Ssc}^* = (\zhat_{ij_1}^*, ..., \zhat_{ij_s}^*)\trans$ and $\zhat_{ij}^* = z_{ij} - \zbar_j(\bx_i)$.

\subsection{Connection to mixed effects models}\label{connections}
In this section, we demonstrate that one can arrive at the quantity \eqref{q0} via a more familiar mixed effects model. Consider the model
\begin{align}\label{mixed_model}
&y_{ir} =  \mu(t_{ir}) + \sum_{k=1}^K\xi_{ik} \phi_k(t_{ir}) + \epsilon_{ir}, \\
&\quad \bxi_i = (\xi_{i1}, ..., \xi_{iK})\trans \sim N(\bB\bz^*_{i\Ssc}, \Lambda), \quad \epsilon_{ir} \sim N(0, \sigma^2)
\end{align}
where $\bB$ is a $K\times s$ matrix with $(k,j)$th entry $\beta_{kj}$ and $\Lambda = \text{diag}(\lambda_1, ..., \lambda_K)$.
We can obtain $Q_0$ as the variance component score test statistic
for $H_0: \bB = 0$. Specifically, let $\beta_{kj} = \eta\nu_{kj}$ and we consider a working model such that $\{\nu_{kj}\}$ are independently distributed with $E(\nu_{kj})=0$ and
$\var(\nu_{kj}) = \lambda_k^2$.
Under this working model, $H_0: \bB =0$ is equivalent to
\[
H_0: \eta = 0.
\]
This formulation follows the logic of variance component score tests that have been proposed previously \cite{wu2011rare} and recalls, for example, the likelihood ratio test proposed in \cite{Crainiceanu2004}. To obtain the variance component test statistic, rewrite the model as \[
\by_{\mu i} = \sum_{k=1}^K\left(\sum_{j\in\Ssc}\eta\nu_{kj}z^*_{ij} + e_{ik}\right)\bphi_{ik} + \bepsilon_i
\]
for centered outcome $\by_{\mu i} = \{y_{i1} - \mu(t_{i1}), ..., y_{ir_i} - \mu(t_{ir_i})\}\trans$, error vector $\bepsilon_i = (\epsilon_{i1}, ..., \epsilon_{ir_i})\trans$, and random effects $\be_i = (e_{i1}, ..., e_{iK})\trans \sim N(0, \Lambda)$. Then
\[
\left.\by_{\mu i}\right| \bnu, \{\bz_{i\Ssc}^*\}_{i=1}^n \sim N\left(\sum_{j\in\Ssc}\sum_{k=1}^K\eta\nu_{kj}z^*_{ij}\bphi_{ik}, \Sigma_{\by_i}\right)
\]
where $\Sigma_{\by_i} = \sum_{k=1}^K \lambda_k\bphi_{ik}\bphi_{ik}\trans + \sigma^2I_{r_i}$ and $I_{r_i}$ is the $r_i \times r_i$ identity matrix.

 The log-likelihood for $\by_{\mu i}$ can then be written 
\begin{align*}
\log\Lsc(\eta) =
-\frac{1}{2} \sumin\left\lbrace\log\left|\Sigma_{\by_i}\right| + 
\left(\by_{\mu i} - \eta\sum_{j\in\Ssc}\sum_{k=1}^K\nu_{kj}z^*_{ij}\bphi_{ik}\right)\trans \Sigma_{\by_i}\inv \left(\by_{\mu i}  - \eta\sum_{j\in\Ssc}\sum_{k=1}^K\nu_{kj}z^*_{ij}\bphi_{ik}\right)\right\rbrace .
\end{align*}
Because the target of inference is $\eta$, we marginalize over the nuisance parameter $\bnu$ conditional on the observed data to obtain
$\Lsc^*(\eta) = E\{\Lsc(\eta) | \mathbb{V}\}$ where the expectation is taken over the distribution of $\bnu$. We follow the argument in \cite{commenges1995score} and note that the score at the null value is 0: $\lim_{\eta \to 0}\partial \log\Lsc^*(\eta) / \partial \eta = E\left(\sumin \by_{\mu i}\trans \Sigma_{\by_i}\inv \sum_{j\in\Ssc}\sum_{k=1}^K\nu_{kj}z^*_{ij}\bphi_{ik} \mid \Vbb\right) = 0$. So we instead consider the score with respect to $\eta^2, \lim_{\eta \to 0} \partial \log \Lsc^*(\eta) / \partial (\eta^2)$, and we show in the supplementary article \cite{supp}
that it can be written
\begin{align*}
&E\left\{\left.\frac{\partial \log\Lsc(\eta)}{\partial \eta}\right|_{\eta = 0} \mid \Vbb\right\}^2 + E\left\{\left.\frac{\partial^2 \log\Lsc(\eta)}{\partial \eta^2}\right|_{\eta = 0} \mid \Vbb\right\}\\
&= E\left(\sumin \by_{\mu i}\trans \Sigma_{\by_i}\inv \sum_{j\in\Ssc}\sum_{k=1}^K\nu_{kj}z^*_{ij}\bphi_{ik} \mid \Vbb\right)^2 - 
E\left(\sumin  \sum_{j, j'\in\Ssc}\sum_{k,k' = 1}^K\nu_{kj}z^*_{ij}\bphi_{ik}\trans \Sigma_{\by_i}\inv \nu_{k'j'}z^*_{ij'}\bphi_{ik'} \mid \Vbb\right)\\
%&= E\left[\sum_{j\in\Ssc}\sum_{k=1}^K\nu_{kj}\left(\sumin\by_{\mu i}\trans \Sigma_{\by_i}\inv z^*_{ij}\bphi_{ik}\right)\mid \Vbb\right]^2 - 
% E\left[\sum_{j\in\Ssc}\sum_{k=1}^K\nu_{kj}^2\left\lbrace\sumin  (z^*_{ij})^2\bphi_{ik}\trans \Sigma_{\by_i}\inv \bphi_{ik}\right\rbrace\mid \Vbb\right]\\
&= \sum_{j\in\Ssc}\sum_{k=1}^K\left(\sumin\by_{\mu i}\trans \Sigma_{\by_i}\inv \bphi_{ik}\lambda_kz^*_{ij}\right)^2 - 
\sum_{j\in\Ssc}\sum_{k=1}^K\left\lbrace\sumin  (\lambda_k z^*_{ij})^2\bphi_{ik}\trans \Sigma_{\by_i}\inv \bphi_{ik}\right\rbrace
\end{align*}
up to a scaling constant.

To finally obtain $Q_0$, we standardize by $n\inv$ and drop the second term because it converges to a constant, yielding the score statistic \begin{align*}
&n\inv\sum_{j\in\Ssc}\sum_{k=1}^K\left(\sumin\by_{\mu i}\trans \Sigma_{\by_i}\inv \bphi_{ik}\lambda_kz^*_{ij}\right)^2 
= n\inv\sum_{j\in\Ssc}\sum_{k=1}^K\left(\sumin\xitilde_{ik}z^*_{ij}\right)^2 = Q_0,
\end{align*}
taking note of the form of $\xitilde_{ik}$ from \eqref{xitilde}. Thus, our proposed test statistic can be obtained as a variance component test under a normal mixed model framework. On the other hand, we can also view $Q_0$ as a simple summary of the overall covariance between the scores of the FPCA and the genetic markers. We next derive the null distribution of the FPVC test statistic without requiring the normal mixed model to hold.

\subsection{Estimating the null distribution of the test statistic}\label{s:diet}
In order to obtain p-values for FPVC testing, we must identify the null distribution of $Q$. To this end, we show in \cite{supp} that the key quantity in $Q$
$$q_{kj} = \nnhalf\sumin\xihat_{ik}\zhat_{ij}^*$$ is asymptotically equivalent to
$$\qtilde_{kj} = \nnhalf\sumin\xitilde_{ik}\zhat_{ij}^*$$
under $H_0$, i.e. $q_{kj} - \qtilde_{kj} = o_p(1)$ for each $j$ and $k$. The key idea for deriving the null distribution of $q_{kj}$ is that, since $\zhat_{ij}^*$ is approximately mean 0 conditional on $\bx_i$, the variability due to approximating $\xitilde_{ik}$ by $\xihat_{ik}$ does not contribute any additional noise to $q_{kj}$ (compared to $\qtilde_{kj}$) at the first order under $H_0$. Thus, we can obtain the limiting distribution of $Q$ by analyzing the quantity $\Qtilde = \sum_{j \in \Ssc}\sum_{k=1}^K \qtilde_{kj}^2. $

To characterize the null distribution of $\Qtilde$, we need to account for the variability in the estimated model parameters for $\mu_{z_j}(\bx_i) = g_j(\btheta_j,\bx_i)$ in $\zhat_{ij}^*$. Without loss of generality, we assume that for each $j$
\begin{equation}
\nhalf(\bthetahat_j - \btheta_j) %= \nnhalf\sumin \Usc_j(\bx_i)\{z_{ij} - g_j(\btheta_j,\bx_i)\}+ o_p(1)
= \nnhalf\sumin \Usc_j(\bx_i) z_{ij}^*+ o_p(1),  \label{exp-thetahat}
\end{equation}
where $ \Usc(\cdot)$ is some ($q+1$)-dimensional function of $\bx_i$ with $E\{\Usc(\bx_i)^2\} < \infty$. It follows that
\begin{align}
q_{kj} = \qtilde_{kj} + o_p(1) = \nnhalf\sumin \Qsc_{ikj}+ o_p(1)
\label{u}
\end{align}
where $\Qsc_{ikj} = \left\{\xitilde_{ik} - \Abb_{kj}\Usc(\bx_i)\right\}z^*_{ij}$,
$\Abb_{kj} = E\{\xitilde_{ik}\dot{\bg}_j(\btheta_j,\bx_i)\trans\}$, and $\dot{\bg}_j(\btheta_j,\bx_i) = \partial g_j(\btheta_j,\bx_i)/\partial \btheta_j$. We show in %the appendix that
\cite{supp} that
%Web Appendix C that
%
the limiting null distribution of $Q$ is a mixture of $\chi^2_1$ random variables, $Q \sim \sum_{l =1}^{sK} a_l\chi^2_1$, with mixing coefficients determined by the eigenvalues of the covariance matrix of $\{\Qsc_{ikj}\}_{j \in \Ssc, 1\leq k \leq K}$. So finally we obtain a p-value for the association between the set $\bz_\Ssc$ and $Y(\cdot)$ as
$P(\sum_{l =1}^{sK} \ahat_l\chi^2_1 > Q \mid \Vbb)$, where $\ahat_l$ is an empirical estimate of $a_l$.

By a similar argument, one could construct an asymptotically equivalent test statistic by estimating $\bxitilde_i$ in two stages. Instead of obtaining an estimator directly from FPCA via equation \eqref{xihat}, FPCA can be used to estimate only $\mu(\cdot)$ and $\{\phi_k(\cdot)\}_{k=1}^K$. By plugging the estimated $\muhat(\cdot)$ and $\{\phihat_k(\cdot)\}_{k=1}^K$ into the mixed model \eqref{mixed_model}, one can obtain what we will call the {\em re-fitted} test statistic
\begin{align}\label{qbar}
\Qbar = n\inv\sum_{j\in\Ssc}\sum_{k=1}^K \left[\sum_{i=1}^n\bar{\xi}_{ik}\zhat^*_{ij}\right]^2
\end{align}
where $\bar{\bxi}_i = (\bar{\xi}_{i1}, ..., \bar{\xi}_{iK})\trans$ is the BLUP from the model $y_{ir} - \muhat(t_{ir}) =  \sum_{k=1}^K\xi_{ik} \phihat_k(t_{ir}) + \epsilon_{ir}$ with $\cov(\bxi_i) = D$, for some unspecified positive definite  matrix $D$.
By the same argument above, estimation of $\xitilde_{ik}$ by $\bar{\xi}_{ik}$ contributes no additional variability to the test statistic at the first order. It follows that  \[
q_{kj}^\dag = \nnhalf \sumin \bar{\xi}_{ik}\zhat_{ij}^* = \qtilde_{kj} + o_p(1).
\]
and hence $\Qbar$ has the same limiting null distribution as $Q$. Not surprisingly, simulation results suggest that the performance of $\Qbar$ is quite similar to the performance of $Q$.
This equivalence indicates that effectively our proposed testing procedure uses FPCA to estimate potentially nonlinear bases and assesses the effect of genetic markers by fitting a mixed model with these basis functions. On the other hand, the test statistics can also be viewed as a simple summary of covariances, and -- since we estimate the null distribution without relying on the normality assumption required by the mixed models -- our testing procedure remains valid regardless of the adequacy of the mixed model.

\subsection{Combining multiple sources of outcome information}
In the HIV progression study, we seek to test the overall association between SNPs and both lCD4 and lVL simultaneously because more and distinct information about HIV progression is captured in both measures than in either one alone. FPVC testing, as outlined above, can be easily adapted to
perform a test for the overall association between $\bz_\Ssc$ and all outcomes of interest.
To use information in multiple outcomes, $\{\by\sm\}_{m=1}^M$, we simply perform FPCA separately on each $\by\sm$ and obtain FPCA scores for each person and each outcome. Subject $i$'s scores for $\by_i\sm$ would be $\bxihat_i\sm =(\xihat_{i1}\sm, ..., \xihat\sm_{iK_m})\trans$, as in \eqref{xihat}, and the full set of scores for person $i$ would be $\bxihat_i = (\bxihat_i^{(1)\transpose}, ..., \bxihat_i^{(m)\transpose})\trans$. Then we simply proceed by testing \[
H_0: \{\by\sm\}_{m=1}^M \perp \bz_\Ssc \mid \bx
\] as before based on
\begin{align}\label{multiple}
Q = \| \nnhalf\sumin \bxihat_i \bzhat_{i\Ssc}^{*\transpose}\|_F^2 =
\sum_{m=1}^M \| \nnhalf\sumin \bxihat_i^{(m)} \bzhat_{i\Ssc}^{*\transpose}\|_F^2.
\end{align}
Since each outcome may be measured on a different scale, one may use scaling or weighting to allow scores from each outcome to contribute similarly to the test statistic. See section \ref{discussion} for further discussion of scaling/weighting.

\section{Association study for HIV progression}\label{s:eg}

In this study, two independent cohorts were recruited in Botswana to detect sets of SNPs related to HIV disease progression as measured by lCD4 and lVL. The first cohort, which we will denote BHP010, was a natural history observational prospective cohort study recruited from clinics in Gaborone. 
This cohort included HIV-1C-infected individuals with CD4 cell counts above 400 cells per $\mu$l and not yet qualified for the Botswana highly active antiretroviral treatment (HAART) program. Patients were not enrolled if they were younger than 18, had an active AIDS-defining illness requiring the initiation of HAART, presented with an AIDS-related malignancy, or had previously been exposed to HAART during pregnancy or breast feeding. 

Follow-up visits occurred at approximately three-month intervals with an additional visit one month after enrollment. VL was generally collected at six-month intervals, and most patients in this cohort do not have VL measurements after two years of follow-up. Follow-up began in 2005 and lasted for up to 255 weeks. The mean follow-up time was 41 months. 
%The end state of the study was administrative censoring (216/456), or enrollment on HAART treatment (121/456), enrollment in other studies (28/456), or loss to follow-up, defined as missing three consecutive visits (91/456). 
At least two CD4 measurements were required for measuring disease progression, and 449 patients satisfied this criterion. Of these, the median age at baseline was 34 years old with an interquartile range (IQR) of (28, 39). There were 366 women. In 2008, 143 patients were genotyped on an Illumina LCG BeadChip, the chip used in the second cohort. After making exclusions for quality control -- call rate greater than 0.99, gender matching listed gender -- 137 individuals were included in the association study.

The second cohort we will denote BHP011. This cohort came from a randomized, multifactorial, double-blind placebo-controlled trial conducted between December 2004 and July 2009 \cite{baum2013effect}. The purpose of the trial was to determine the efficacy of micronutrient supplementation (supplementation of multivitamin, selenium, or both) in improving immune function in HIV-1C-infected individuals. It was composed of 878 treatment-naive patients with CD4 higher than 350 cells/$\mu$l, as well as body mass index (BMI) greater than 18 for women and 18.5 for men (calculated as weight in kilograms divided by height in meters squared), age of 18 years or older, no current AIDS-defining conditions or history of AIDS-defining conditions, and no history of endocrine or psychiatric disorders.

Patients were followed up for a maximum of 169 weeks. They returned to clinics approximately every 3 months to measure CD4 and approximately every 6 months to measure VL. The mean follow-up time was 696 days. Of these, 838 had at least two CD4 measurements, of whom 613 were women. The median age at baseline was 33 years old with an interquartile range (IQR) of (28, 39). In this cohort, 326 individuals were genotyped on Illumina LCG BeadChips, with 320 entered into the association study after quality control exclusions.

FPCA was performed on each cohort separately for both lCD4 and lVL. Three eigenfunctions were chosen for lCD4 and two for lVL in each cohort, which corresponds to 99\% of proportion of variance explained for each. The form of the eigenfunctions look similar for both lCD4 and lVL in each cohort and lend themselves to reasonable interpretations. The first eigenfunction tends to serve as a mean shift or an intercept; the second eigenfunction acts something like a slope; and the third eigenfunction behaves approximately as a quadratic term. The vector of estimated "re-fitted" scores (refer to \eqref{qbar}) for each individual to be used in testing can be written $\bxihat_i = (\xihat_{ik}\sm)_{k=1,2,3; m = 1,2}$ where $\xihat_{ik}\sm$ is the estimated score corresponding to the $k$th estimated eigenfunction of lCD4 when $m = 1$ and lVL when $m=2$. %
%\begin{figure}
%\centering
%\begin{tabular}{c}
%\includegraphics[width = 0.45\textwidth, keepaspectratio = true]{chr6_mhtn_plot.png}
%\end{tabular}\vspace{0.2in}
%\caption{.}\label{trajs}
%\end{figure}

A total of 155,007 SNPs on chromosome 6 were genotyped. After requiring less than 5\% missingness and at least five individuals with any minor alleles in each cohort, $p = 108,665$ SNPs, $\bz_i = (z_{i1}, ..., z_{ip})$ remained for association testing, where the ordering in $\bz_i$ corresponds to position on the chromosome. The dominant model was used for analysis such that $z_{ij} = 1$ if any minor alleles are present and $z_{ij}$ = 0 if none are present. Missing values were imputed as the minor allele frequency for that SNP. In order to gain power by pooling information in nearby SNPs, sets of 10 contiguous SNPs were constructed as $\bz_{i\Ssc_j} = (z_{ij}, ..., z_{ij+9})$ for $j = 1, ..., 108,656$. Here the choice of 10 merely serves as example for illustration and sets could in principle be constructed with more SNPs, but due to the small sample size in each cohort, we kept the size of the sets modest. Tests were performed on each cohort separately, and p-values were combined using the Fisher method. The false discovery rate was controlled at 0.1 using the Benjamini-Hochberg procedure \cite{benjamini1995controlling}, which is expected to remain valid since although the moving window construction of sets induces high correlations for nearby regions, SNP sets in distant regions are not expected to be correlated \cite{storey2004strong}. 

Tests were adjusted for age and gender to remove any possible confounding, so that we are testing for the effect of SNP sets on disease progression conditional on age and gender. Logistic regression was used to remove their effect. The method appears to be robust to this specification, as results were not markedly changed either when no adjustment was made or by specifying a probit model (results not shown). 

\begin{figure}
\centering
\begin{tabular}{c}
\includegraphics[width = 0.45\textwidth, keepaspectratio = true]{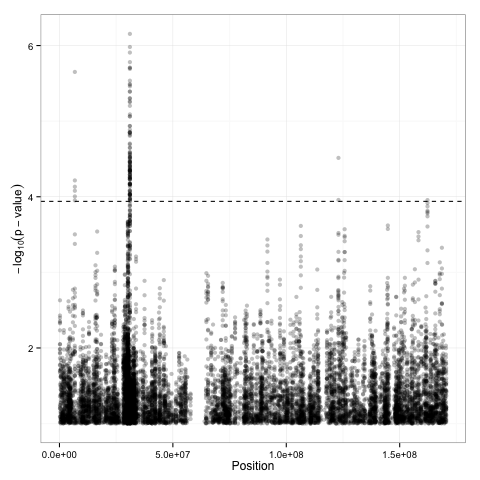}
\end{tabular}\vspace{0.2in}
\caption{Manhattan plot for set-based testing on chromosome 6. Position on x-axis for each test is determined by the middle SNP (5th of 10)	 in the set. The dotted line corresponds to the threshold for rejection at FDR 0.1.}\label{mhtn}
\end{figure}

The manhattan plot for the 108,656 tests is given in figure \ref{mhtn}. 
In all, 126 tests passed the FDR threshold, corresponding to 4 broad regions of the chromosome. Six contiguous tests rejected in the region between positions 6,784,416 and 6,793,116 (region 1), which fall between the LY86 and BTF3P7 genes; 117 tests rejected between positions 31,022,266 and 31,080,899 (region 2), including SNPs on the HCG22 and C6orf15 genes; two tests rejected between 122,990,817 and 123,014,708 (region 3) on the PKIB gene; and one test rejected representing SNPs in the region between 162,250,522 and 162,254,546 (region 4) including SNPs on the PARK2 gene. Notably, the C6orf15 gene has been reported to be associated with susceptibility to follicular lymphoma \cite{skibola2009genetic}, and genes in linkage disequilibrium with HCG22 and C6orf15 have demonstrated associations to total white blood cell counts \cite{nalls2011multiple} and multiple myeloma \cite{chubb2013common}.

\begin{figure}
\centering
\begin{tabular}{cc}
(a) \includegraphics[width = 0.45\textwidth, keepaspectratio = true]{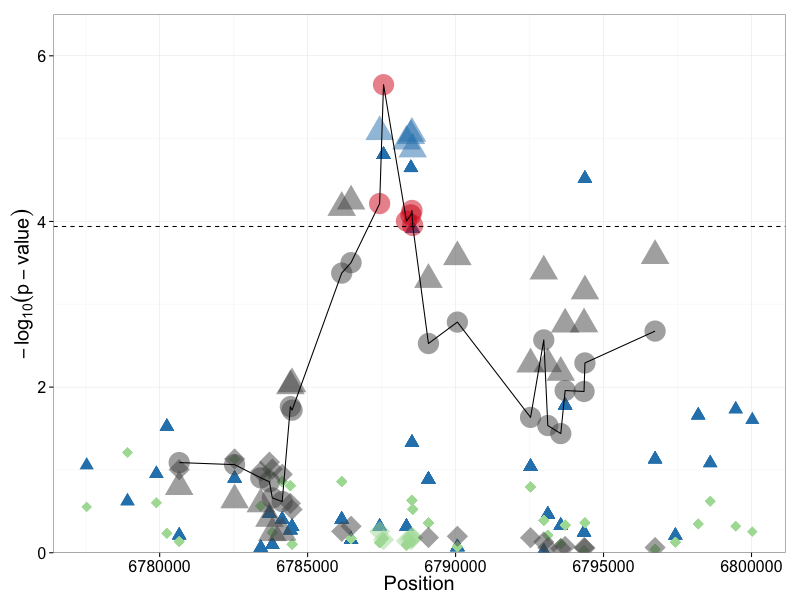} & (b) \includegraphics[width = 0.45\textwidth, keepaspectratio = true]{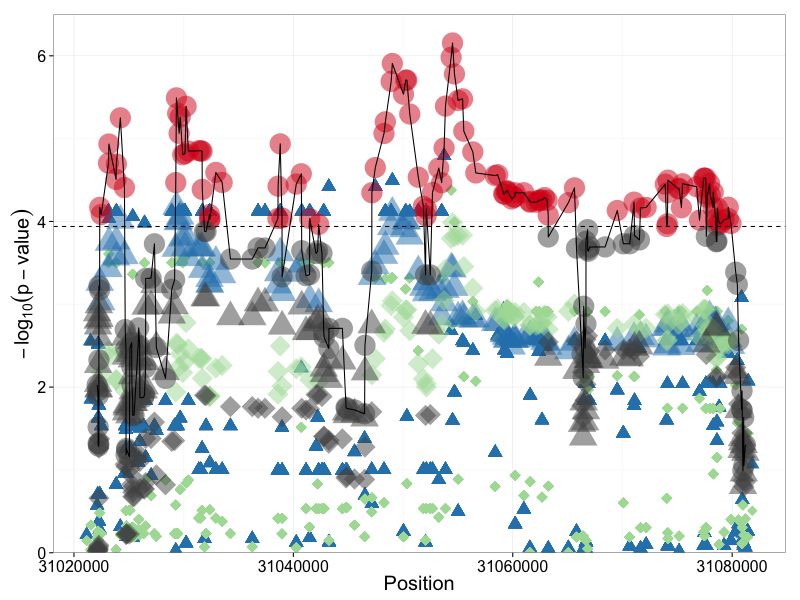}\\
(c) \includegraphics[width = 0.45\textwidth, keepaspectratio = true]{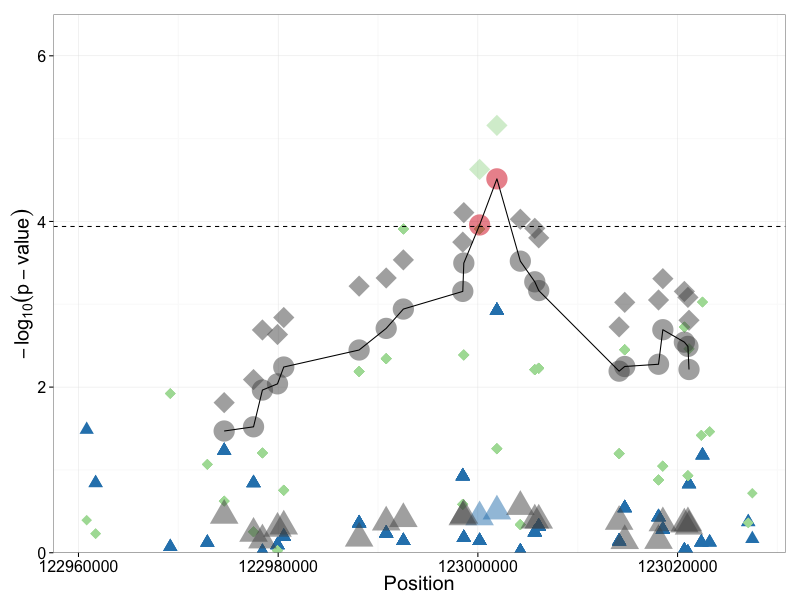} & (d) \includegraphics[width = 0.45\textwidth, keepaspectratio = true]{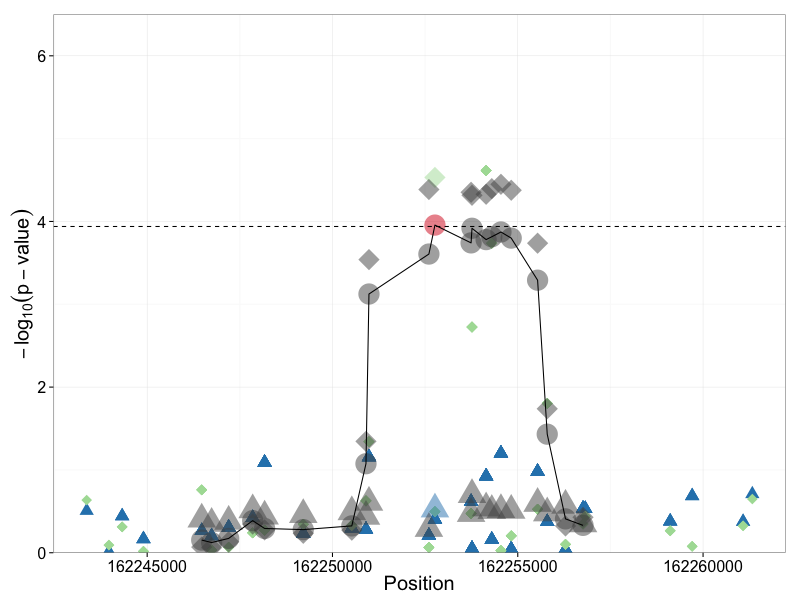}\\
\end{tabular}\vspace{0.2in}
\caption{P-values in significant regions on the -log$_{10}$ scale. Large symbols correspond to set-based tests, and for illustration small symbols correspond to tests for individual SNPs. Triangles represent p-values computed in BHP010, and diamonds BHP011. Circles represent combined set-based p-values, which are of primary interest and are connected by lines. Combined p-values that are below the FDR threshold are in color, as are their corresponding component p-values.}\label{regs}
\end{figure}

Furthermore, regions 1, 3, and 4 -- whose -log$_{10}$ p-values are depicted in figure \ref{regs} (a), (c), and (d), respectively -- only have strong signals in one of the two cohorts. Region 1 demonstrates association largely in BHP010, as can be seen in the figure where the large triangles in the figure (representing the set-based p-value in BHP010) tend to lie above the large dots (representing the combined set-based p-value), while the diamonds for BHP011 tend to be very low. Conversely, in regions 3 and 4 the association is apparent only in BHP011. Whereas in region 2, associations tend to be strong in both cohorts, and the combined p-values tend to be lower (higher in the figure) than either of the component p-values.

\begin{figure}
\centering
\begin{tabular}{cc}
(a) \includegraphics[width = 0.4\textwidth, keepaspectratio = true]{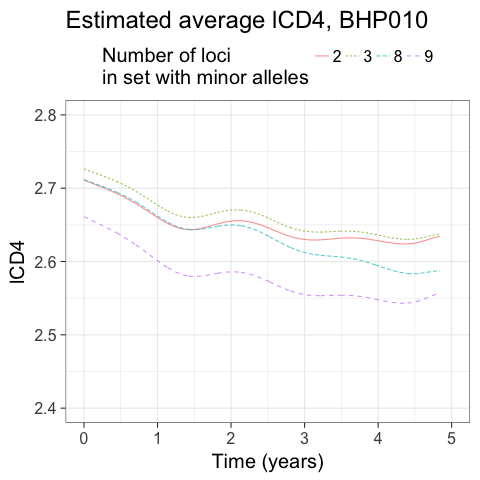} & (b) \includegraphics[width = 0.4\textwidth, keepaspectratio = true]{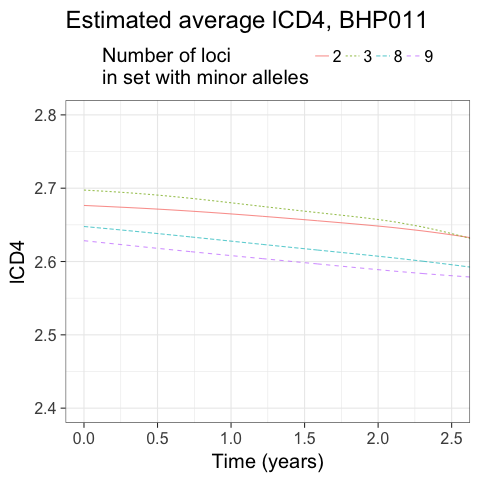} \\
(c)\includegraphics[width = 0.4\textwidth, keepaspectratio = true]{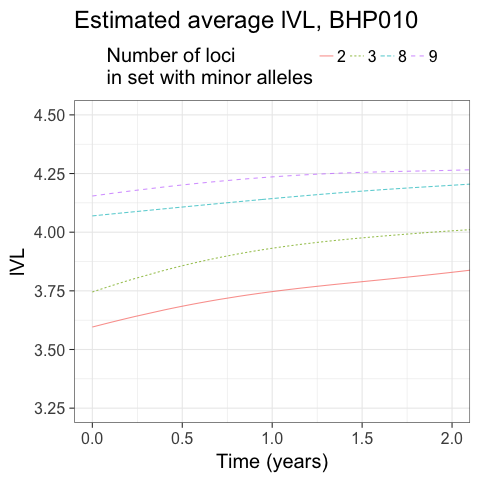} & (d) \includegraphics[width = 0.4\textwidth, keepaspectratio = true]{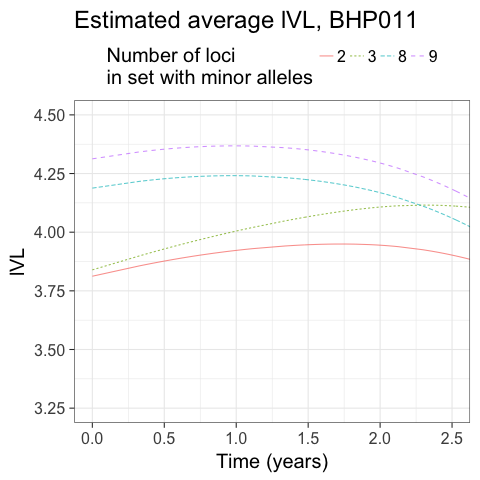}\\
\end{tabular}\vspace{0.2in}
\caption{Disease progression by minor allele burden in most significant SNP set. Estimated lCD4 and lVL are grouped by number of loci in SNP set with any minor alleles and averaged. For clarity, just those individuals with 2, 3, 8, and 9 loci are included. Lines correspond to estimates of the conditional mean based on FPCA.}\label{smths}
\end{figure}

To get a better sense of what the test is picking up on, we looked at average disease progression within groups of patients with similar minor allele burden. To do this, we identified the SNP set with the smallest p-value, which lay in the HCG22 portion of region 2 and included the following SNPs: rs2535308, rs2535307, rs2535306, rs2535305, rs3130955, rs2535304, rs12527394, rs2535303, kgp9442190, and rs3130959. Patients were grouped according to the number of loci among these 10 at which they had any minor alleles. Within these groups, we averaged the estimated mean lCD4 and lVL in each cohort over time $\Yhat(\cdot) = \muhat(\cdot) + \sum_{k=1}^K\xihat_{ik}\phihat_k(\cdot)$. As a demonstration, the results for those with 2, 3, 8, and 9 loci with minor alleles are depicted in figure \ref{smths}. We selected these groups to demonstrate burden extremes (very few individuals had 0 or 1 loci affected, so they were not shown). 

Those with only 2 and 3 affected loci looked to generally be the healthiest group, having higher and more stable CD4 and lower and gently increasing VL throughout the study period in each cohort. Those with 3 affected loci tended to have a more negative CD4 slope and higher and increasing VL over the study period. Those with 9 affected loci in general had the worst progression: low and declining CD4 counts in both cohorts, and high and relatively stable VL in both cohorts. Those with 8 affected loci tend to fall in the middle. Smoothing the raw data directly in each of these groups yielded similar results and nearly identical conclusions (results not shown).

\section{Simulation results}\label{simulation}
We have performed simulation studies to assess the finite sample performance of our proposed testing procedure and compare its power to the standard
linear-mixed-model-based procedures. For simplicity, we focused on a single marker $z$ in the absence of covariates and two potential functional outcomes generated from
\begin{align*}
y_{ir}^{(m)} &= Y_i\sm(t_{ir}) + \epsilon_{ir}\sm\\
&=  \sin(t_{ir}) + (-1)^{m-1} \gamma\left\{\sin(t_{ir}/3) + \cos(t_{ir})\right\} + 
(1-\gamma)\left\{ b_{i0} + 0.5b_{i1}^{(m)}\cos(t_{ir}/4)\right \} + \\
&\qquad \beta z_i\left\lbrace \alpha\left(\cos(t_{ir}) + \cos(t_{ir}/10) - \sin(3t_{ir})\right)\right. + 
\left.(1-\alpha)t_{ir}/7\right\rbrace + \epsilon_{ir}^{(m)}, \qquad m = 1, 2,
\end{align*}
where $b_{ij}^{(m)} \sim N(0,0.25), j = 0, 1$ are independent and identically distributed (iid) random effects and $\epsilon_{ir}^{(m)} \sim N(0,0.25)$ are iid errors, for $m = 1, 2$. For each subject $i$, we generate the number of observations from a Poisson distribution $r_i \sim$ Poisson($\lambda$) + 2, and we generate $t_{ir}$ uniformly over the time interval $(0, 2\pi)$. The parameter $\beta$ controls the magnitude of the genetic effect. The parameter $\alpha$ controls how linear the genetic affect is -- when $\alpha = 0$ the genetic effect is entirely linear, and when $\alpha = 1$ the effect is entirely nonlinear. On the other hand, $\gamma$ controls the complexity of the mean process and the amount of inter-subject variability -- when $\gamma = 0$, the mean process is relatively simple but the inter-subject variability is high, and when $\gamma = 1$ the mean process is complex and the inter-subject variability is low. The genetic factor $z_i$ is generated according to a binomial(2, 0.1), with 0.1 the minor allele frequency.

We examined the performance of the FPVC test statistic $Q$ (defined in \eqref{multiple}, here denoted by ``FPCA") and its asymptotically equivalent counterpart $\Qbar$ (defined in the context of a single outcome in \eqref{qbar}, here denoted "Re-fitted"). For the purposes of comparison, we also examined the performance of a similar test statistic that does not use FPVC but instead employs a pre-specified basis. Consider the test statistic
$Q_{\text{lin}} = \frac{1}{n}\sum_{m=1}^2\sum_{k=1}^2 \left[\sum_{i=1}^n\xi^{(m)\dag}_{ik}\zhat^*_i\right]^2$, where $\xi^{(m)\dag}_{ik}$ is the BLUP from the linear mixed model $y_{ir} =  \beta_0 + \beta_1t_{ir} + \xi_{i1} + \xi_{i2}t_{ir} + \epsilon_{ir}$. In the following, we denote results for $Q_{\text{lin}}$ by "Linear".

The number of FPCA scores for the $m$th outcome, $K_m$, was selected as the smallest $K$ such that
the fraction of variation explained (FVE),  $\sum_{k=1}^K \lambdahat_k / (\sum_{k} \lambdahat_k)$,
was at least $\wp = 0.99$. To ensure that the scores for each outcome contributed comparably to the test statistics, we centered and scaled each outcome as $y_{ir}\smstar = (y_{ir}\sm - \ybar\sm)/\sigmahat_y\sm$, prior to obtaining $\xihat\sm_{ik}$ and $\xi_{ik}^{(m)\dag}$, where $\sigmahat\sm_y = \sqrt{(n-1)\inv\sum_{i,r}(y\sm_{ir} - \ybar\sm)^2}$ and $\ybar\sm = n\inv\sum_{i,r} y\sm_{ir}$.

In the following we report power as the proportion of 1000 simulations for which the testing procedure produced a p-value below 0.05 to demonstrate the relative performance of the various testing procedures. To ensure that the asymptotic null distribution of the test statistic yields a valid testing procedure, we evaluate the entire distribution of p-values under the null hypothesis, including at levels much lower than 0.05.

\subsection{Type I error}
\begin{figure}
\centering
\begin{tabular}{c}
\includegraphics[width = 0.4\textwidth, keepaspectratio = true]{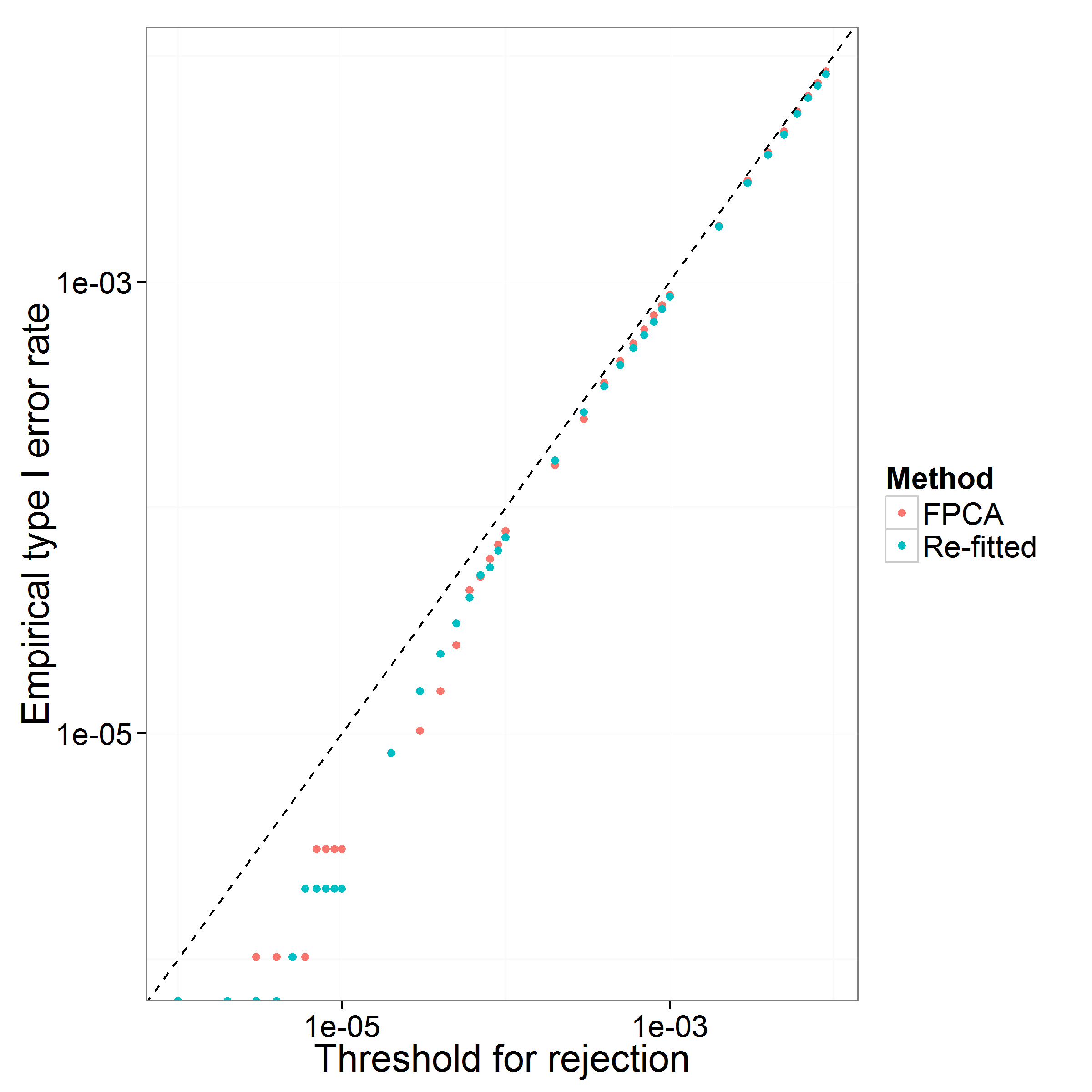}
\end{tabular}\vspace{0.2in}
\caption{Empirical type I error rates for tests performed at various levels, based on $10^6$ simulations.}\label{null_p_plot}
\end{figure}

In the following we take $\lambda = 6$. The empirical type I error rates for testing at the 0.05 level ranged from 0.040 ($\gamma$ = 0.25) to 0.048 ($\gamma$ = 0) for $Q$; from 0.036 ($\gamma$ = 1) to 0.047 ($\gamma$ = 0) for $\Qbar$; and from 0.040 ($\gamma$ = 0.75) to 0.059 ($\gamma$ = 1) for $Q_{\text{lin}}$. However, levels much smaller than 0.05 are necessary to control error rates in large-scale testing. Thus, in order to establish the validity of our testing procedure for performing many tests, we establish that the resulting p-values are approximately uniform under the null hypothesis. We performed $10^6$ simulations under the null, with $n = 200, \gamma = 1,$ and $\alpha = 0$, and we obtained the type I error of FPVC testing at each of the following levels: $1\times10^{-6}, ..., 9\times 10^{-6}, 1\times10^{-5}, ..., 9\times 10^{-5}, 1\times10^{-4}, ..., 9\times 10^{-4}, 1\times10^{-3}, ..., 9\times 10^{-3}$. Results are depicted in Figure \ref{null_p_plot}. The Figure shows that the level is preserved at all levels of testing. Further simulations would provide better approximations of type I error rates at smaller levels, but these results suggest that the asymptotic null distribution fits quite well in small samples.

\subsection{Power}\label{s:pwr}
\begin{figure}
\centering
\begin{tabular}{c}
\includegraphics[width = \textwidth, keepaspectratio = true]{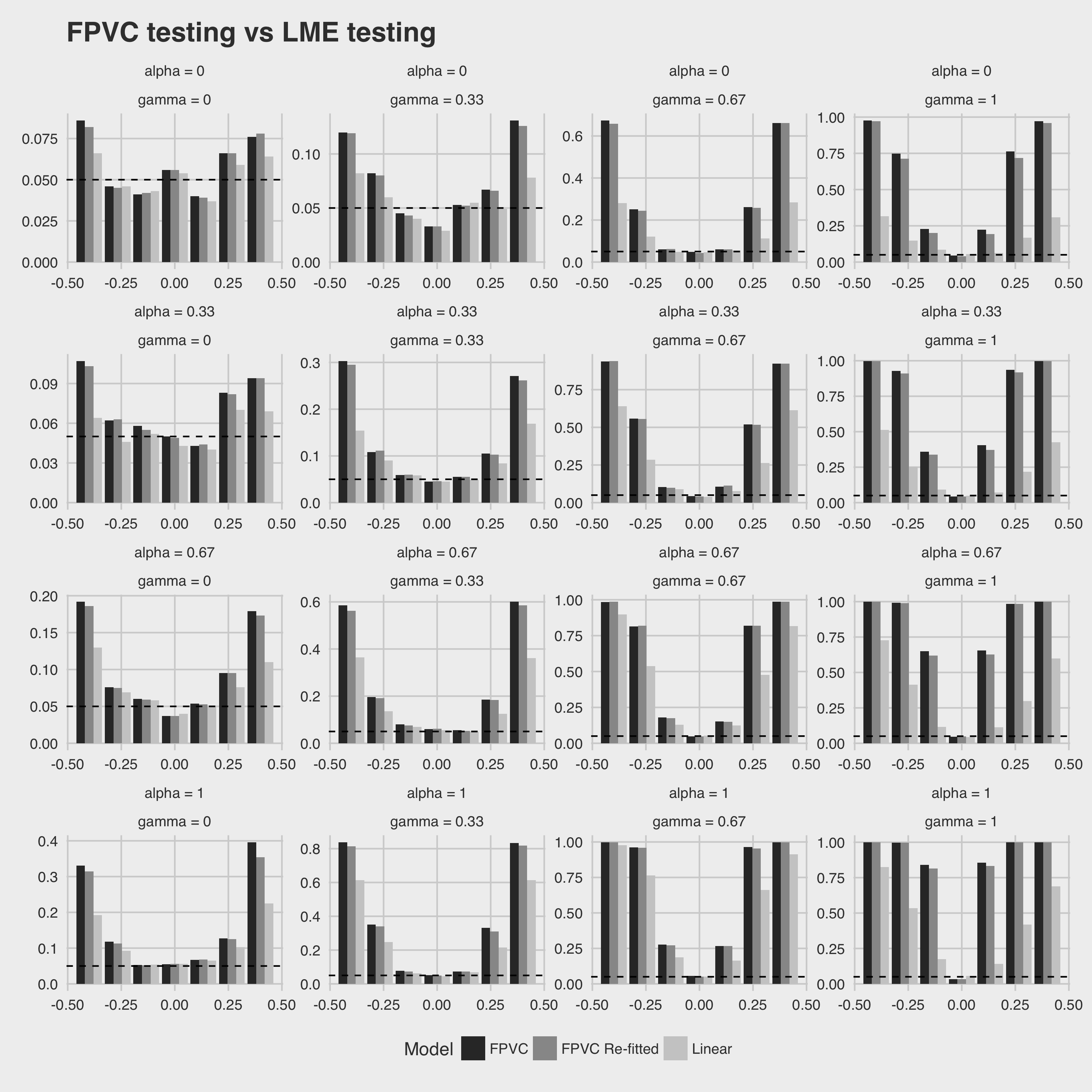}
\end{tabular}\vspace{0.2in}
\caption{Power to detect $\beta$ using $Q$ (FPVC), $\Qbar$ (FPVC Re-fitted), and $Q_{\text{lin}}$ (Linear).}\label{linear_power}
\end{figure}
%The Figure indicates that the performances of $Q$ and $\Qbar$ was similar and uniformly dominated the performance of $Q_{\text{lin}}$ at any sample size and MAF. %Each test maintained its nominal size, with empirical type I error rates for $Q$ and $\Qbar$ ranging between 0.034 (MAF = 0.2, $n$ = 400) and 0.057 (MAF = 0.2 and $n$ = 600 for $Q$ and MAF = 0.1 and $n$ = 400 for $\Qbar$) and for $Q_{\text{lin}}$ between 0.036 (MAF = 0.2 and $n$ = 400) and 0.055 (MAF = 0.1 and $n$ = 600).

In figure \ref{linear_power}, we display results for $n = 100$ and all levels of $\gamma$ and $\alpha$. There, the figure demonstrates that, despite the fact that the true effect was linear when $\alpha = 0$, both FPC-based tests dominate the linear-based tests, and the advantage of using FPVC, in terms of power, did not disappear, even when the true effect was linear. %Again, the empirical type I error rates hovered around the desired value of 0.05: ranging from 0.041 ($\alpha$ = 0.75) to 0.057 ($\alpha$ = 0.25) for $Q$; from 0.037 ($\alpha$ = 0.75) to 0.048 ($\alpha$ = 0) for $\Qbar$; and from 0.046 ($\alpha$ = 0.75) to 0.065 ($\alpha$ = 0.25) for $Q_{\text{lin}}$.
Notably, as $\gamma$ varied, we saw power gains by using the FPVC-based $Q$ and $\Qbar$, with the gains increasing as $\gamma$ approached 1 and the functional form of $Y_i\sm(\cdot)$ became more complex, and the need to flexibly model it increased. 

In all of our simulations, the FPVC methods dominated the linear method in terms of power while maintaining desirable type I error rates. We wanted to ensure that the improvement we were seeing was not simply due to the fact that the linear model used only two scores, a random intercept $\xi^{(m)\dag}_{i1}$ and a random slope $\xi^{(m)\dag}_{i2}$, for each outcome whereas the FPVC-based methods used $K_m$ scores, where $K_m$ was often selected larger than 2. Thus, we also considered the performance of scores based on fixed-basis expansions of $\bt$, using either polynomial or spline bases.

%Specifically, we compared the FPVC test statistics $Q$ and $\Qbar$ with $K_m = 2$ to $Q_{\text{lin}}$; we compared the FPVC test statistics with $K_m = 3$ to $Q_{\text{quad}} = \frac{1}{n}\sum_{m=1}^2\sum_{k=1}^3 \left[\sum_{i=1}^n\xi^{(m)\dag}_{ik}\zhat^*_i\right]^2$, where $\xi^{(m)\dag}_{ik}$ is the BLUP from the mixed model $y_{ir} =  \beta_0 + \beta_1t_{ir} + \beta_2t_{ir}^2+ \xi_{i1} + \xi_{i2}t_{ir} + \xi_{i3}t_{ir}^2+ \epsilon_{ir}$; and we compared the FPVC test statistics with $K_m = 4$ to $Q_{\text{cube}} = \frac{1}{n}\sum_{m=1}^2\sum_{k=1}^4 \left[\sum_{i=1}^n\xi^{(m)\dag}_{ik}\zhat^*_i\right]^2$, where $\xi^{(m)\dag}_{ik}$ is the BLUP from the mixed model $y_{ir} =  \beta_0 + \beta_1t_{ir} + \beta_2t_{ir}^2+ \beta_3t_{ir}^3 + \xi_{i1} + \xi_{i2}t_{ir} + \xi_{i3}t_{ir}^2+ \xi_{i4}t_{ir}^3 + \epsilon_{ir}$.

Specifically, we fit models with $K = 2, 3, ..., 6$ degrees of freedom. For polynomial bases corresponded to the model $y_{ir} = \sum_{k=1}^K(\beta_k + \xi_{ik})\ttilde_{ir}^{k-1} + \epsilon_{ir}$ for $\ttilde_{ir}$ a centered and scaled version of $t_{ir}$. And for the spline basis, we used cubic B-splines constructed with the specified degrees of freedom with the \texttt{bs} function in the \texttt{splines} R package. Because, in some sense, FPCA does model selection by choosing the basis that explains the most variability in $\by$, we also perform model selection on the pre-specified bases in order to make the comparison fair. We select the model with the lowest AIC and use the $\xihat_{ik}$s from that model in the testing procedure. We will call the test statistic based on B-splines $Q_B$ and the model based on polynomial bases $Q_p$.

\begin{figure}
\centering
\begin{tabular}{c}
\includegraphics[width = \textwidth, keepaspectratio = true]{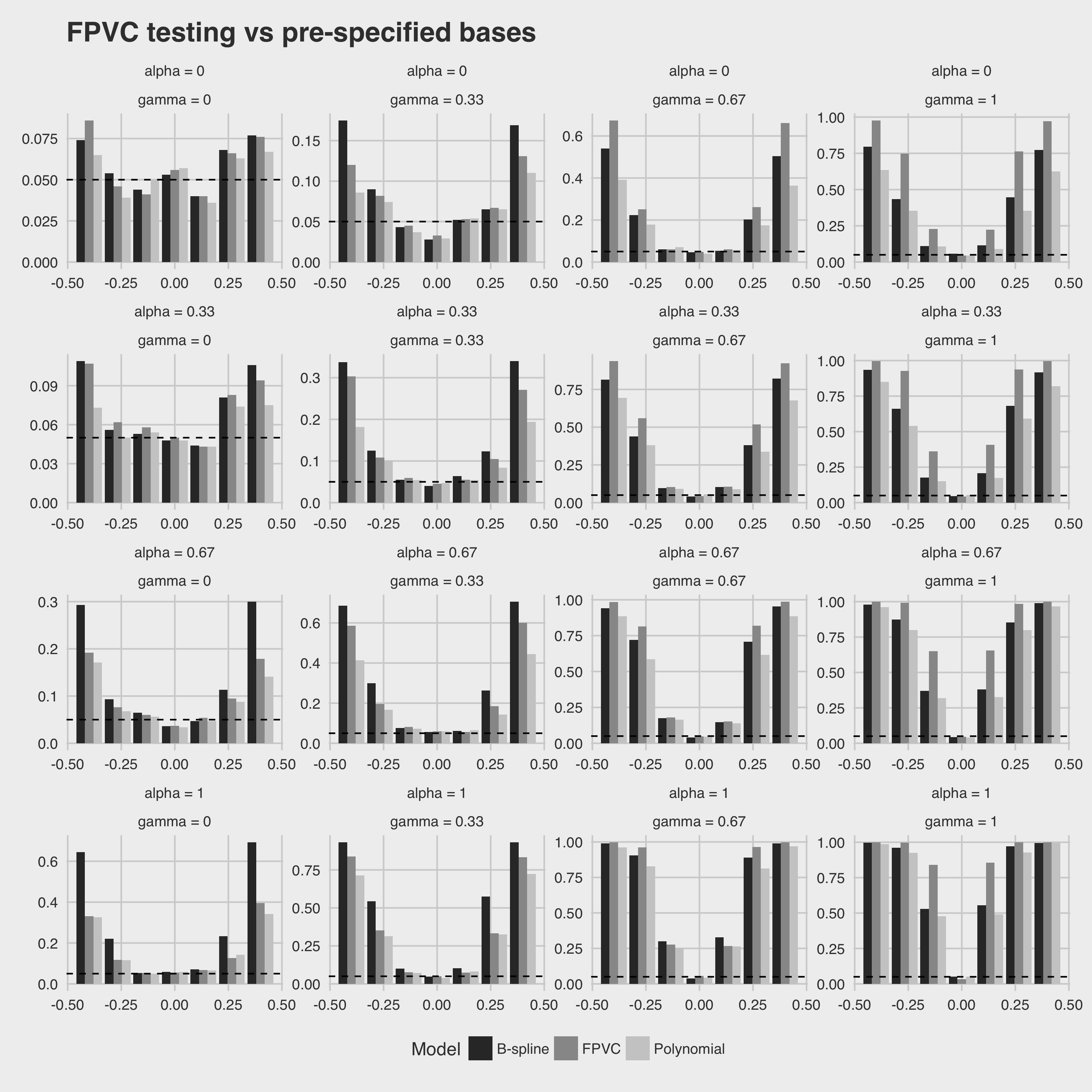}
\end{tabular}\vspace{0.2in}
\caption{Power to detect $\beta$ using $Q$ (FPVC),  $Q_B$ (B-splines), and $Q_p$ (Polynomial).}\label{mod_power}
\end{figure}

Results are found in figure \ref{mod_power}. We found that there were some situations when using the pre-specified B-spline basis could outperform the FPVC tests, particularly when $\gamma$ was near 0 (low mean complexity) and $\alpha$ was near 1 (linear genetic effect). However, the polynomial basis never outperformed FPVC. And as the complexity of the trajectory $\gamma$ increased, the desirability of FPVC testing always increased. This suggests that in simple problems, using a pre-specified basis may be preferable, but for complex effects and complex trajectories, FPVC will likely be preferred. And in general, if the complexity of the trajectory is unknown, FPVC testing offers a generally powerful method for all settings that is insensitive to tuning parameter selection.

\section{Discussion}\label{discussion}
We have proposed functional principal variance component testing, a FPCA-based testing procedure for assessing the association between a set of genetic variants $\bz_\Ssc$ and a complexly varying longitudinal outcome $\by$ that is feasible on the genome-wide scale, allowing adjustment for other covariates. Unlike the standard mixed-model-based approaches, we do not model the trajectories $\{Y_i(\cdot)\}_{i=1}^n$ parametrically but use the data to identify the most parsimonious summaries of the trajectory patterns via FPCA. We subsequently test the association between the random coefficients $\bxi_i$ and the markers of interest using a test statistic motivated by variance component testing. Our procedure could potentially be much more powerful than procedures based on pre-specified bases, which might suffer power loss due to either high degrees of freedom or inability to capture the complexity in the trajectories.
Furthermore, our FPVC testing is computationally efficient as we are able to perform thousands or even millions of tests quickly by separating the time-intensive FPCA from the testing. This makes our method feasible on the genome-wide scale where millions of marginal tests may be necessary. As an example, computing test statistics and p-values for FPVC testing typically takes less than 0.1 seconds for a set of 10 SNPs and both lCD4 and lVL combined. On the other hand, fitting a single linear mixed effects model for only lCD4 with a random effect for a small pre-specified B-spline basis takes more than 2 seconds. At the genome-wide scale we would observe a speed-up on the order of hours. Code for FPVC testing is available at \texttt{https://github.com/denisagniel/fpvc}.

It is important to note that while we make mild assumptions on the longitudinal outcome $\by$ to obtain the form of our proposed test statistic, the validity of FPVC testing requires no assumption about the relationship between $\by$ and $\bz_\Ssc$. FPVC testing remains valid even if the working mixed model \eqref{mixed_model} fails to hold.
Additionally, while one can motivate the quantity $\xitilde_{ik}$ as the conditional expectation of $\xi_{ik}$ under a normality assumption on $\xi_{ik}$ and $\epsilon_{ir}$, even when this normality fails to hold, testing based on $Q$ remains valid since the estimated eigenvalues and eigenfunctions from functional PCA are uniformly convergent to their limits \cite{hall2006properties}. In fact, one can consider FPVC model-free in that the test statistic $Q$ could be motivated simply as an estimated covariance. Furthermore, we assume that the errors $\epsilon_{ir}$ are iid with mean 0 and variance $\sigma^2$, but some relaxation of this assumption is possible for some "degree of weak dependence and in cases of non-identical distribution" \cite{hall2006properties}, while still maintaining the validity of our procedure.

FPVC testing can also simultaneously consider multiple sources of outcome information to better characterize complex phenotypes. With multiple longitudinal outcomes, one might wish to ensure that scores for all outcomes are roughly on the same scale, so that each outcome contributes  comparably to the test statistic. To this end,
one may consider a weighted version of \eqref{multiple} as \[
Q = \sum_{m=1}^M \omega_m \| \nnhalf\sumin \bxihat_i \bzhat_{i\Ssc}^*\|_F^2,
\] where $\omega_m$ are nonnegative outcome-specific weights that can be pre-specified or
data adaptive. Alternatively, in the absence of relevant weights, one can simply scale each $\by\sm$ so that the magnitude of $\bxihat_i\sm$ is comparable across different values of $m$. Let $y\smstar_{ir} = y\sm_{ir}/\sigmahat\sm_y$ where $\sigmahat\sm_y = \sqrt{(n-1)\inv\sum_{i,r}(y\sm_{ir} - \ybar\sm)^2}$ and $\ybar\sm = n\inv\sum_{i,r} y\sm_{ir}$. Then obtain $\bxihat_i\smstar$ via FPCA on $\{\by_i\smstar\}_{i=1}^n$ and construct the test statistic $\sum_{m=1}^M \| \nnhalf\sumin \bxihat_i\smstar \bzhat_{i\Ssc}^*\|_F^2$. Such a strategy appears to work well in simulation studies.

While we use FPCA to summarize the longitudinal trajectories for the purpose of testing with low degrees of freedom, in principle another suitably parsimonious nonparametric method could be used instead. For example, if observations were measured on a common, fine grid of points, then one could imagine using the methods in \cite{morris2006wavelet} to first regress $\by$ on $\bt$, obtain the random effects estimates (similar to the $\bxi_i$ employed here), and use these in testing. However, no available approaches are as widely applicable as our FPCA-based approach, which can be used even when data are sparsely observed; other approaches may not have as small effective degrees of freedom as an FPCA-based method, and the resulting testing procedure may be more sensitive to correct tuning.

% \begin{supplement}[id=suppA]
%   \sname{Supplement A}
%   \stitle{Supplementary assumptions, proofs, and plots}
%   \slink[doi]{COMPLETED BY THE TYPESETTER}
%   \sdatatype{.pdf}
%   \sdescription{We provide additional proofs to support the main theoretical findings of the paper, as well as additional plots to demonstrate testing performance in finite samples.}
%   \end{supplement}

\bibliography{bibtwo}
\bibliographystyle{plain}

\end{document}